\PassOptionsToPackage{dvipsnames}{xcolor}
\documentclass[sigconf, authorversion, nonacm]{acmart}

\AtBeginDocument{%
  \providecommand\BibTeX{{%
    \normalfont B\kern-0.5em{\scshape i\kern-0.25em b}\kern-0.8em\TeX}}}

\usepackage{lineno}

\usepackage{algorithm}
\usepackage{algpseudocode}
\usepackage[normalem]{ulem}
\usepackage{hhline}
\usepackage{graphicx}
\usepackage{caption} 
\usepackage{float}
\usepackage{numprint}
\usepackage{placeins}
\usepackage{lipsum}
\usepackage{atbegshi}
\usepackage{multicol}

\begin{document}

\title{Multi-Resolution Diffusion for Privacy-Sensitive Recommender Systems} 



\author{Derek Lilienthal}
\affiliation{%
 \institution{San Jose State University}
 \streetaddress{1 Washington Sq, San Jose, CA 95192}
 \city{San Jose}
 \country{United States}}
\email{derek.lilienthal@sjsu.edu}

\author{Paul Mello}
\affiliation{%
 \institution{San Jose State University}
  \streetaddress{1 Washington Sq, San Jose, CA 95192}
  \city{San Jose}
  \country{United States}}
\email{paul.mello@sjsu.edu }

\author{Magdalini Eirinaki}
\affiliation{%
 \institution{San Jose State University}
  \streetaddress{1 Washington Sq, San Jose, CA 95192}
  \city{San Jose}
  \country{United States}}
\email{magdalini.eirinaki@sjsu.edu}

\author{Stas Tiomkin}
\affiliation{%
 \institution{San Jose State University}
 \streetaddress{1 Washington Sq, San Jose, CA 95192}
 \city{San Jose}
 \country{United States}}
\email{stas.tiomkin@sjsu.edu}

\renewcommand{\shortauthors}{Lilienthal et al.}

\begin{abstract}

While recommender systems have become an integral component of the Web experience, their heavy reliance on user data raises privacy and security concerns. Substituting user data with synthetic data can address these concerns, but accurately replicating these real-world datasets has been a notoriously challenging problem. Recent advancements in generative AI have demonstrated the impressive capabilities of diffusion models in generating realistic data across various domains. In this work we introduce a Score-based Diffusion Recommendation Module (SDRM), which captures the intricate patterns of real-world datasets required for training highly accurate recommender systems. SDRM allows for the generation of synthetic data that can replace existing datasets to preserve user privacy, or augment existing datasets to address excessive data sparsity. Our method outperforms competing baselines such as generative adversarial networks, variational autoencoders, and recently proposed diffusion models in synthesizing various datasets to replace or augment the original data by an average improvement of 4.30\% in Recall@$k$ and 4.65\% in NDCG@$k$.

\end{abstract}


\received{31 October 2023}

\maketitle

\textit{This work has been submitted to the IEEE for possible publication. Copyright may be transferred without notice, after which this version may no longer be accessible.}

\section{Introduction}

Recommender systems have become a ubiquitous part of the Web experience, with applications in multiple domains. Such systems are trained on user data that come in the form of explicit ratings or implicitly inferred preferences for items as well as various other sources of information including demographics, item metadata, context data, social network connections, etc. \cite{springerRecommenderSystems}. Inevitably, this heavy reliance on user data makes privacy one of the major challenges for recommender systems \cite{Paraschakis16, MilanoTF20}. Breaches of user privacy and even violations of rights may occur in various stages of a recommender system's life cycle, including inferences that can be made from the output of the recommendation process \cite{Friedman15, Paraschakis16, MilanoTF20}. Research into privacy-preserving algorithms and anonymization of user data has made significant progress, but a balance between accurately capturing complex user preferences and data privacy remains challenging \cite{Privacy_Preserving_Gen_Data_Rec_Sys}.
Even when data privacy is not the primary concern, there are other needs which can be addressed by synthetic datasets for recommender systems, such as data sparsity \cite{IdrissiZ20}. A dataset is considered sparse when a very small percentage of total user-item preferences are available. In cases where a high number of missing values exists, a synthetic dataset can be used to augment the original dataset by increasing the total number of data points, ideally improving the prediction accuracy of the recommender system.

Generative AI has become increasingly popular due to its ability to generate realistic content across various modalities \cite{Harshvardhan20, Bond-TaylorLLW22}. 
Generative models such as Generative Adversarial Networks (GANs) \cite{Goodfellow14}, Variational Autoencoders (VAEs) \cite{Kingma2014}, and Diffusion models (DMs) \cite{ho2020denoising} have been shown to model complex data distributions extremely well, and have thus been employed in various domains, including recommender systems. Most prior work employs GANs or VAEs as part of the collaborative filtering process (i.e. to model user-item interactions) \cite{AugCF, liang2018variational, Askari21, CFGAN, UGAN, ma2019learning, sequential_VAE}, with the objective of improving predictions rather than generating synthetic data. DMs, a recent addition to the generative modeling family, have emerged as a prominent framework for data synthesis and have been shown to produce high-quality samples, outperforming GAN architectures \cite{dhariwal2021diffusion}. Recently, DMs have been applied to recommendation systems for collaborative filtering \cite{Walker2022, Diffusion_Recommender_Model} and sequential predictions \cite{li2023diffurec, sequential_recommendation_diffusion_model} with very promising results. 

In this work, we propose a Score-based Diffusion Recommendation Module (SDRM) that allows the generation of synthetic data in the form of user-item interactions. SDRM utilizes a VAE encoding scheme to map user-item preferences to a latent Gaussian distribution, where a diffusion model transforms the Gaussian input and reconstructs it using the VAE decoder. The training of our encoder, decoder, and diffusion model are conducted separately, which allows us to leverage the VAE's variational inference capabilities. Through this architecture, we can generate arbitrarily many synthetic samples that accurately capture the users' preference distribution allowing for a partial or total replacement of the user data. 

This approach results in a module that can produce diverse data while preserving the complex distributions of user-item interactions. 
Through thorough experimental evaluation, our approach has been proven to be more effective than other methods that utilize GANs \cite{xu2019modeling}, VAEs \cite{xu2019modeling, liang2018variational}, and DMs \cite{Walker2022, Diffusion_Recommender_Model} in creating recommendation datasets of varying sizes and densities. 
We demonstrate SDRM's ability to generate accurate synthetic recommendation data across four datasets and three collaborative filtering recommender algorithms, showcasing the performance increase when using SDRM compared to the original data and other data generation techniques. 

Our contributions are summarized as follows\footnote{We have made our code available at \url{https://github.com/Multi-resolution-diffusion-recommender/SDRM}}:
\begin{itemize}
    \item  A score-based diffusion module (SDRM) for generating high quality synthetic data from user-item preference datasets.
    \item  A novel method to boost the accuracy of a collaborative filtering algorithm through variational inference with a score-based diffusion model and variational autoencoder.
    \item The introduction of multi-resolution sampling in the diffusion process to further increase the accuracy of synthetic data-based recommendations. 
    \item An extensive experimental evaluation across four datasets and three recommendation algorithms demonstrates increased accuracy by an average improvement of 6.81\% in Recall and 7.73\% in NDCG for augmented datasets, and 1.42\% in Recall and 1.98\% in NDCG for synthetic datasets—over competing methods. This is achieved while preserving privacy by attaining 99\% dissimilarity to the original data and maintaining similar user and item distributions.
\end{itemize}

To the best of our knowledge, this is the first work that employs diffusion models with the primary objective to generate synthetic datasets for recommender systems.

The rest of this paper is organized as follows: Section 2 provides an overview of related work in privacy methods for recommender systems, synthetic data generation using generative models, and generative modeling for recommender systems. Section 3 provides a background on the core materials used in our approach, namely VAEs and DMs. In Section 4, we present SDRM, detailing our architectural choices, including the proposed objective function and the sampling and training methodologies. Section 5 presents the setup and results of our experimental evaluation across a range of datasets and evaluation metrics. Finally, in Section 6, we summarize our results, discuss our module's limitations, and outline our plans for future work.

\section{Related Work}

In this section, we review related work and how it relates to ours from three different angles: privacy in recommender systems, synthetic data generation, and generative models for recommender systems.

\begin{figure*}
    \centering
    \caption{SDRM Training and Sampling}
    \includegraphics[trim={0 5cm 0 5cm},clip,width=\linewidth, height=140pt]{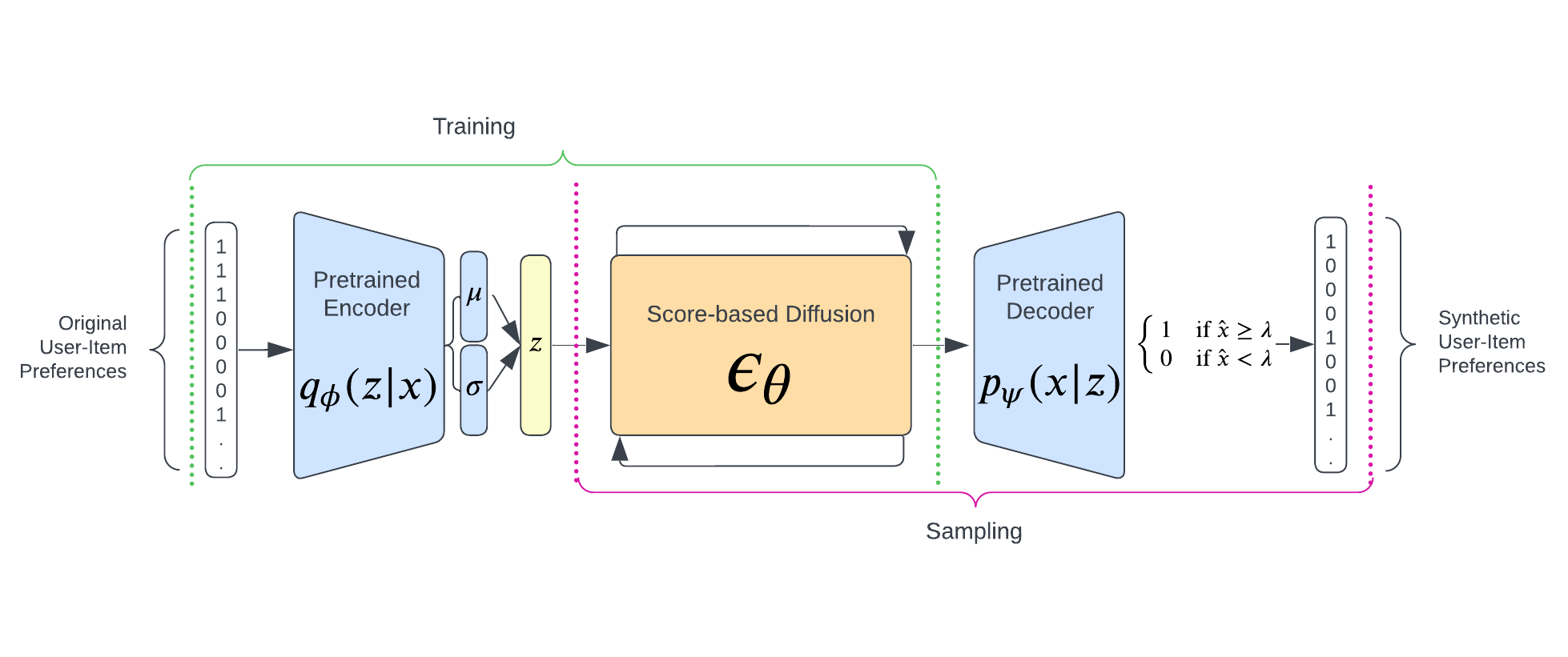}
    \label{fig:SDRM_figure}
\end{figure*} 

\subsection{Privacy in Recommendations}

Traditional methods of privatizing data involved removing personally identified information from the datasets, therefore anonymizing the data. However, such methods have been shown to be ineffective as large-scale datasets could easily be de-anonymized \cite{narayanan2007robust}. This led to the formulation of differential privacy, which mathematically guarantees that the data or a model trained with differentially private data is private \cite{foundations_of_dp}.

There exist several works that focus on differentially private recommender systems employing matrix factorization \cite{pd_matrix_fact} or federated learning approaches to decentralize a users' data from a single location \cite{asad2023comprehensive}. While differential privacy methods provide a mathematical guarantee the data is private, it incurs a decrease in accuracy when compared to training the model with the original dataset \cite{bagdasaryan2019differential}. To counteract this loss in performance, prior works have been proposed that focus on synthesizing data using 'privacy-preserving' techniques such as Liu et al.'s UPC-SDG \cite{Privacy_Preserving_Gen_Data_Rec_Sys} and Slokom et al.'s CART-based methods \cite{slokom2020partially}. While such techniques do not mathematically guarantee private data, they aim to enhance user privacy without compromising accuracy. UPC-SDG does so through an attention-based module substituting user preferences, while Slokom et al. use partially generated data for privacy. While our objective is the same, in our approach we focus on augmenting or replacing original datasets with synthetic data using a score-based diffusion method for recommender algorithm training.

\subsection{Synthetic Data Generation}

Many prior works using diffusion have focused on improving image classification through augmentation. These methods have focused on generating more training examples to improve the performance of downstream tasks, like building more robust classifiers \cite{azizi2023synthetic, trabucco2023effective} or improving image segmenting for healthcare applications \cite{Conditional_Diffusion_Model-Based_Data_Augmentation}. 

The need for fully synthetic copies of datasets has been driven by the need of companies to stay compliant with data protection regulations \cite{Lesnikowski2021SyntheticDA}, such as the GDPR (General Data Protection Regulation) \cite{EuropeanParliament2016a}. Many works have been proposed to create data generators whose objective consists of synthesizing tabular datasets which hold the same statistical properties as the original, but are not identical copies \cite{Lei19, tabbDDPM, ctgan_plusp}. Both CTGAN \cite{xu2019modeling} and CTAB-GAN+ \cite{ctgan_plusp} use Generative Adversarial Networks, while for TabDDPM \cite{ho2020denoising} the authors use a Denoising Diffusion Probabilistic Model (DDPM) \cite{ho2020denoising} to model tabular datasets. However, none of these prior works have benchmarked their models on recommender system datasets, leaving a gap in the current research for examining the generalizability of tabular data generators to model recommendation datasets.

Our method, SDRM, utilizes a diffusion process similar to TabDDPM for modeling recommendation data, which could be broadly categorized as tabular. However, SDRM distinguishes itself by compressing categorical features (user-items preferences) into lower-dimensional representations via a Variational Autoencoder to address the high dimensionality of recommendation data and employs multi-resolution sampling for new data generation. In contrast, TabDDPM's approach of concatenating each categorical feature as the input into the denoising MLP (Multilayer Perceptron) network significantly increases the computational complexity due to the thousands of unique categories (items) in recommendation datasets. TabDDPM's complexity made direct comparisons between SDRM infeasible in our evaluation in section 5.

\subsection{Generative Models in Recommendations}

Recently, generative models have been used in recommender systems, with most of the prior work involving GANs and VAEs. GAN-based techniques leverage adversarial training to optimize the generator's ability to augment user interactions \cite{gao2021recommender, AugCF}. The application of GANs to enhance user-item data for training recommender models has gained significant attention, with methods emerging to augment user preferences for collaborative filtering \cite{CFGAN, AugCF, UGAN}, and sequential predictions \cite{RecGAN, PLASTIC}. Various works on incorporating GANs and reinforcement learning to simulate user behavior have also been explored \cite{chen2019generative, bai2019model}.

Many works have been proposed to leverage VAEs in recommendation systems to predict user ratings \cite{liang2018variational, ma2019learning, sequential_VAE, Askari21} by using their ability to compress data into a Gaussian distribution using the re-parameterization trick \cite{Kingma2014} and applying variational inference to reconstruct the original data. To take advantage of VAE's abilities to reconstruct data from a Gaussian distribution, we propose pretraining a VAE to map data to a low-dimensional Gaussian representation for which we apply diffusion on their latent space. Similar work has been proposed using this method \cite{vahdat2021score, Diffusion_Recommender_Model}, but trained the VAE and diffusion model together, whereas we train them individually to take advantage of the variational inference ability of VAEs. 

There have been very few works incorporating diffusion into recommendation systems. Existing research primarily focuses on predicting complete user preferences \cite{Walker2022, Diffusion_Recommender_Model} or making sequential predictions \cite{li2023diffurec, sequential_recommendation_diffusion_model}. Walker et al. proposed CODIGEM \cite{Walker2022}, a DDPM-inspired architecture that stacks autoencoders to denoise the corrupted user-item preferences. Very recently, Wang et al. proposed DiffRec \cite{Diffusion_Recommender_Model}, a model that trains a denoising diffusion model by incorporating an array of MultiVAE encoders to map user data to a low-dimensional space and reconstructs the input back to its original high-dimensional form. While DiffRec and CODIGEM show promise as methods to predict user ratings, they are not suitable as data generators and do not take full advantage of diffusion's ability to generate realistic data. We make this claim for a few reasons: first, both DiffRec's and CODIGEM's training approach implements early stopping, which prevents the denoising model from fully learning the entire resolution of the latent data during the denoising process \cite{ho2020denoising}; and second, CODIGEM and DiffRec train for 3 and 10 timesteps $T$ respectively, while training for larger $T$ (such as 1000 steps or more), has been shown to capture more detailed information, due to increased reconstruction time to learn imperceptible details of the data, as demonstrated in \cite{ho2020denoising, pmlr-v139-nichol21a}.

In contrast to DiffRec and CODIGEM, our model, SDRM, benefits from a simpler architecture and trains over a larger amount of \(T\) steps without early stopping and utilizes an ablated training objective which focuses on generating user-item interaction data for enhancing, or completely replacing the original datasets. 
Additionally, we specifically designed SDRM to create synthetic user preferences that closely follow the intricate distribution of user-item preference data. This design enhances the performance of out-of-the-box recommender models while preserving user privacy, rather than serving as a direct predictor model.

\section{Background}

\subsection{Variational Autoencoders}

Variational Autoencoders (VAEs) are a probabilistic generative model that aims to learn a joint distribution between the data and a compressed latent distribution, $p_{\psi}(x, z)$. VAEs use a stochastic encoder network $q_{\phi}(z | x)$, to encode data to a lower-dimensional latent space $z$, then reconstruct it using a stochastic decoder network $p_{\psi}(x | z)$ using variational inference to approximate the probabilistic latent variables. The latent distributions are typically a normal Gaussian, characterized through its mean $\mu$ and variance $\sigma$.

VAEs learn the stochastic mappings between the observed data and the latent variables through the \textit{evidence lower bound} (ELBO) optimization objective. This objective aims to approximate the log-likelihood of the observed data, which is intractable due to the integral of the marginal likelihood, $p_{\psi} = \int p_{\psi}(x,z)dz$, lacking an analytical solution \cite{intro_to_vaes}. The ELBO objective is defined as follows:
\begin{equation}
\begin{aligned}[t]
    L(x ; \psi, \phi) & \equiv \mathbb{E}_{q_\phi (z | x)} \left[ \log p_\psi (x | z) \right] \\ 
                        & - KL\left(q_\phi(z | x) \parallel p(z)\right) 
\end{aligned}
\end{equation}

We can obtain an unbiased estimation of the ELBO by sampling the latent $z$ space $z \sim q_{\phi}$ and optimizing it through gradient ascent. However, to take gradients with respect to $\phi$ in this sampling process, we must employ the \textit{reparameterization trick}: $z = \mu_{\phi}(x) + \epsilon \odot \sigma_{\phi}(x)$ \cite{Kingma2014}, where $\epsilon \sim \mathcal{N}(0,I_K)$ and $K$ represents the size of the latent space. This trick introduces stochasticity into the sampling procedure, allowing gradients with respect to $\phi$ to be back-propagated through the latent $z$ space.

Using the reparameterization trick, we can optimize the encoder network parameters, $\phi$, and decoder network parameters, $\psi$, using backpropagation while maintaining the stochasticity needed to model an unbiased latent distribution (such as Gaussian) during the training process. To create new data samples, we need to sample some random noise from a normal Gaussian distribution $\epsilon \sim \mathcal{N}(0,I_{K})$ and pass it to the decoder network $p_{\psi}(x | \epsilon)$ to obtain a new data sample $\hat{x}$ which follows the approximate distribution of an original data sample $x$. 

SDRM leverages the abilities of the decoder network of a VAE to transform Gaussian noise into data samples that approximate the original data distribution. We then model these VAE latent $z$ distributions using a score-based diffusion model. Parameterizing this latent distribution using score-based diffusion allows us to take advantage of diffusion models ability to generate high-quality data samples, which in this case is Gaussian noise at different resolutions, and the variational inference capabilities of VAEs to generate new samples from a lower-dimensional representation. 

\subsection{Diffusion Models}

Diffusion Models (DMs) have emerged as a versatile class of generative models. DMs are probability density functions which consist of two processes, a {\it forward} and a {\it reverse} \cite{sohldickstein2015deep}. In the forward process, noise perturbations are incrementally added to the initial input $x_{0}$, often in the form of a standard normal Gaussian $x_{T}\sim N(0, \mathbf{I})$, using a noise scheduler $\beta_{t}$ for $T$ time steps. This forward process is a conditional distribution $x_{t}$ given $x_{t-1}$ and can be defined as follows:

\begin{equation}
    q(x_{t}|x_{t-1}) := \mathcal{N}(x_{t}; \sqrt{1 - \beta_{t}}x_{t-1}, \beta_{t}I)
\end{equation}

In the reverse process, a neural network $\theta$ learns to predict the next state from the prior state by gradually denoising an input from $x_{T}$ to $x_{0}$. Here, $\mu_{\theta}(x_t, t)$ and $\Sigma_{\theta}(x_t, t)$ defines the mean location and information densities respectively.

\begin{equation}
    p_{\theta}(x_{t-1}|x_t) := \mathcal{N}(x_{t-1}; \mu_{\theta}(x_t, t), \Sigma_{\theta}(x_t, t))
\end{equation}

The reverse process parameters, $\theta$, are maximized through an evidence lower bound (ELBO) objective on the original data distribution \cite{ho2020denoising}. The ELBO in DM\textit{s} can be represented by the mean square error between the original data, corrupted by the forward process, and the reconstruction data, generated by the reverse process. This enables the efficient training of DMs and high-quality data synthesis \cite{ho2020denoising, luo2022understanding, NEURIPS2021_b578f2a5, kingma2023understanding}.

Generative models, such as diffusion, can be trained utilizing various training objectives \cite{Chefer_Attend-and-Excite, zhang2023metadiff}, including a score-based approach \cite{pascap_DSM_connections, song2020generative, song2021scorebased, batzolis2021conditional, luo2022understanding, dockhorn2022scorebased, NEURIPS2022_105112d5, voleti2022scorebased, gnaneshwar2022scorebased, sun2023scorebased}. Compared to the ELBO, the unbiased nature of score-based objectives naturally lends itself to recommender systems. Score-based objective functions, like score matching, are defined by taking the gradient of the log likelihood of $x$ and minimizing the expected squared error of the score based predictions.

\begin{equation}
     \mathbb{E}_{p(x)} \left[ \| s_\theta(x) - \nabla \log p(x) \|^2 \right]
\end{equation}

Score matching objectives can suffer from various difficulties including error estimation. When data is sparse, they may not be able to accurately estimate the score function. Simultaneously, access to ground truth data is ideal to accurately capture the score. Our approach avoids these problems by utilizing a VAE to increase the data density and modifies the score-based objective function by introducing noise through $\Tilde{x} = x + \epsilon$ and transforming the function into a denoising problem. We leverage this score-based objective in our work to capture the intricacies of user-item interactions.

\section{Score-Based Diffusion Recommender Module}

We propose SDRM (Score-based Diffusion Recommender Module), a module designed to generate artificial user-item preferences to augment or replace the dataset it is trained on. 

Diffusion models are capable of modeling complex distributions of arbitrarily many dimensions, but suffer from intractability when datasets have sparse features. As a result, reducing the overall dimensions before the diffusion process can help with modeling accurate reconstructions over all possible items of the dataset. For this reason, our proposed method employs a VAE to map data into a lower-dimensional Gaussian distribution and enables more efficient training for SDRM.

Generating recommendation data poses a unique problem due to the challenge of training a model on datasets containing tens or hundreds of thousands of user preferences. Recommendation data is highly sparse and skewed \cite{springerRecommenderSystems}, making it necessary to reduce the overall dimensions before the diffusion process. Through this method, diffusion can learn to reconstruct all possible items of the dataset. For this reason, our proposed method takes advantage of VAE's ability to map data into a lower-dimensional Gaussian distribution and enables more efficient training for SDRM.

As shown in Figure \ref{fig:SDRM_figure}, SDRM uses a pre-trained encoder to map user ratings into a smaller representation, followed by a Multi-Layer Perceptron (MLP) denoising model, and finally, a decoder to decompress the latent vectors into user ratings. We train SDRM in two steps: first, we pretrain a multinomial likelihood variational autoencoder (MultiVAE) on user ratings, and then we train SDRM with the encoder of the MultiVAE. To generate user ratings, we sample from a normal Gaussian distribution, denoise it using SDRM, and pass the denoised latent vectors into the VAE decoder to reconstruct the data into the same dimensions as the original dataset. 

We propose two sampling variations for SDRM: full-resolution sampling (F-SDRM), and multi-resolution sampling (M-SDRM). F-SDRM denoises pure Gaussian noise starting at timestep \(T \rightarrow 0\) while M-SDRM starts at a random timestep $t \in T$ and denoises to \(0\). Because the latent \(z\) vector of the pretrained encoder maps to a Gaussian distribution, SDRM learns to denoise from a Gaussian distribution \(z\) to a Gaussian distribution \(\hat{z}\). By starting at a random timestep \(t\) instead of the full timestep \(T\) and denoising until we reach timestep \(0\), we reconstruct Gaussian noise at different resolutions into user data for M-SDRM.

\subsection{SDRM objective} 

We propose the following score-based objective for training diffusion in recommender systems. 
\begin{equation}
\small
    \ell(z_t) = 
    \frac{||\bigl(s_{\theta}(\hat{z}_t) - s_{\theta}(z_t)\bigr)- \Delta_{\theta}(z_t)||_2^2 + ||s_{\theta}(z_t) - \Delta_{\theta}(z_t)||_2^2}{||\Delta_{\theta}(z_t)||_2^2}
\label{eq:loss}
\end{equation}
where $t$ is the discrete time step, uniformly sampled from the range $[1, \dots, T]$; $z_{t}$ is the corresponding latent representation of the input data point (e.g., user-item interactions) encoded by the VAE; $\hat{z}_{t}=z_t + \nu$ with $\nu \sim \mathcal{N}(0, \sigma_{\nu}\mathbf{I})$ is its perturbed version by the Gaussian noise, $\eta$; $s_{\theta}(\cdot)$ is the score function in the diffusion model with the parameters $\theta$; and $\epsilon_{\theta}(z_t)$ is the prediction noise at the time $t$ in the diffusion model. Finally, $\Delta_{\theta}(z_t) \doteq \epsilon_{\theta}(z_t) - z_t$ quantifies the difference between the prediction noise, $\epsilon_{\theta}(z_t)$, and the latent representation, $z_t$.
The denominator, $||\Delta_{\theta}(z_t)||_2^2$, serves as a normalizing factor to stabilize the loss function. We show, through experimentation on various datasets, that this score-based objective is more effective than standard training objectives applied in other diffusion models, such as CODIGEM \cite{Walker2022} and DiffRec \cite{Diffusion_Recommender_Model}.

\subsection{SDRM Training and Sampling}

We train MultiVAE \cite{liang2018variational} on the original dataset, using the evidence lower-bound objective defined by 
\begin{equation}
\begin{aligned}[t]
    \mathcal{L}_u(\psi, \phi) = & \mathbb{E}_{q_\phi(z_u | x_u)}[\log p_\psi(x_u | z_u)] \\
    & - \beta \cdot KL(q_\phi(z_u | x_u) \| p(z_u))
\end{aligned}
\end{equation} 
where $\log p_\psi(x_u | z_u)$ is the Gaussian log-likelihood for user $u$, $q_\phi$ is the approximating variational distribution (inference model), and $\beta$ is a heuristic annealing parameter. Next, we utilize the MultiVAE encoder to compress user-item preferences into a smaller latent representation for training SDRM with the loss in Equation~\eqref{eq:loss}. We prevent gradient updates on the VAE model while training SDRM, ensuring SDRM purely learns to transform the VAE latent variable \(z\) to another latent variable \(\hat{z}\).

For sampling data with F-SDRM, we follow the same sampling procedure as Ho et al. \cite{ho2020denoising} and map \(z_{0}\) to the decoder of the MultiVAE. We follow the same process for M-SDRM, but start the denoising from a random \( t \in T\) for each sample. 
Additional details of the training and sampling procedure are included in Algorithms \ref{alg:sdrm_training} and \ref{alg:sdrm_sampling}. 

\begin{algorithm}
\caption{Training SDRM}\label{alg:sdrm_training}
\begin{algorithmic}[1]
\footnotesize
\State Sample candidate hyperparameters $x^{*}$ from \(\chi\) and initialize SDRM
\While{Recall@10 improves}
\For{batch size $\in$ $x^{*}$} 
\Statex \hspace{\algorithmicindent} \hspace{\algorithmicindent} \textit{// Train VAE}
\State \begin{math} \begin{aligned}
     \mathcal{L}_u(\psi, \phi) = & \mathbb{E}_{q_\phi(z_u | x_u)}[\log p_\psi(x_u | z_u)] \\
     & - \beta \cdot KL(q_\phi(z_u | x_u) \| p(z_u))
\end{aligned} \end{math}
\EndFor
\EndWhile
\For{epochs $\in x^{*}$} 
\Statex \hspace{\algorithmicindent} \textit{// Map user-item preferences to \(z\)}
\State    $z_u \leftarrow q_{\phi}(z_u|x_u)$   
\Statex  \hspace{\algorithmicindent} \textit{// Train SDRM}
\State       $\ell(z_t) = \frac{||\bigl(s_{\theta}(\hat{z}_t) - s_{\theta}(z_t)\bigr)- \Delta_{\theta}(z_t)||_2^2 + ||s_{\theta}(z_t) - \Delta_{\theta}(z_t)||_2^2}{||\Delta_{\theta}(z_t)||_2^2}$ 
\EndFor
\end{algorithmic}
\end{algorithm}

\begin{algorithm}
\caption{Sampling SDRM}\label{alg:sdrm_sampling}
\begin{algorithmic}[1]
\footnotesize
\State Select $t$ number of denoising timestep from $[1, T]$
\State Initialize synthetic user-item preference vector with sampled Gaussian noise $z_{T} \sim N(0,I)$
\For{$T$,...,$1$}
    \Statex \hspace{\algorithmicindent} \textit{// Sample Gaussian noise}
    \State $z \sim N(0,I)$ if $t>1$, else $z=0$
    \Statex \hspace{\algorithmicindent} \textit{// Transform Gaussian noise to Gaussian noise}
    \State  $\hat{z}_{t-1}=\frac{1}{\sqrt{\alpha_{t}}}\left ( \hat{z}_{t}-\frac{1-\alpha_{t}}{\sqrt{1-\bar{\alpha}}}\epsilon_{\theta}(\hat{z}_{t},t) \right )+\sigma_{t}z$ 
\EndFor
\Statex \textit{// Reconstruct $\hat{z}$ using VAE decoder to $\hat{x}$}
\State $\hat{x} \leftarrow p_{\psi}(x|\hat{z}) $
\Statex \textit{// Apply piecewise conditional to $\hat{x}$ with threshold $\lambda$}
\State $f(\hat{x})=\begin{cases} 1 & \text{if } \hat{x} \geq \lambda \\ 0 & \text{if } \hat{x} < \lambda \end{cases}$
\end{algorithmic}
\end{algorithm}

\section{Evaluation}

We evaluate SDRM by comparing how well the generated user-item data can augment or replace the original data when used as input for any recommendation algorithm. For this purpose, we demonstrate the ability of SDRM to synthesize recommendation datasets by employing three recommendation algorithms, namely SVD \cite{mnih2007probabilistic}, MLP \cite{covington2016deep}, and NeuMF \cite{he2017neural}, and evaluate the generated top-$k$ recommendations in terms of Recall@$k$ and ranking order, using NDCG@$k$. We compare our model's two variations, namely F-SDRM and M-SDRM, with six generative model baselines on four diverse datasets.

\subsection{Experiment Setting}
\subsubsection{Datasets}
We employ four publicly available and commonly used datasets to evaluate SDRM: Amazon Luxury Beauty (ALB) \cite{ni2019justifying}, Amazon Digital Music (ADM) \cite{ni2019justifying}, MovieLens 100k (ML-100k) \cite{grouplensMovieLens}, and MovieLens 1M (ML-1M) \cite{grouplensMovieLens}. We selected these datasets because they encompass a wide range of items and vary in user/item/rating cardinality and sparsity, as shown in Table \ref{tab:dataset_stats}. Moreover, their relatively small size allowed us to conduct extensive, lengthy optimization runs. Each dataset is pre-processed by converting ratings greater than three to 1 and ratings three or less to 0. We also filtered out low-frequency users and items, so only users who rated at least five items and items with at least five ratings are included in the datasets. 

Each original dataset was split into a training, testing, and validation set using a 70:20:10 split on all the users. Additionally, a small portion of user ratings in the testing set are needed as ground truth users in the training dataset to evaluate the effectiveness of synthetic data. The ratings of those users who are in the training and test set are masked during the evaluation phase. 

\begin{table}[ht]
    \centering
    \footnotesize
    \caption{Dataset Statistics}  
    \begin{tabular}{|l|c|c|c|c|} \hline 
         \textbf{Dataset}&  \textbf{\#Users} &  \textbf{\#Items}&  \textbf{\#Ratings} &\textbf{Sparsity}\\ \hline 
         Amazon Luxury Beauty&  1,344&  729&  15,359& 98.43\%\\ \hline
        Amazon Digital Music&  10,621&  8,582&  108,509&99.88\%\\ \hline
         MovieLens 100k&  938&  1,008&  95,215&89.93\%\\ \hline 
         MovieLens 1M&  3,125&  6,034&  994,338&94.73\%\\ \hline 
    \end{tabular}
    \label{tab:dataset_stats}
\end{table}

\subsubsection{Baseline Generative Models}

We evaluate the ability of SDRM to generate synthetic data against the original dataset (representing the baseline before augmenting or substituting data), and the following generative models: 
\begin{itemize} 
    \item CTGAN \cite{xu2019modeling} (Conditional Tabular Generative Adversarial Network) is a GAN designed for synthesizing tabular data. 
    \item TVAE \cite{xu2019modeling} is a conditional Tabular Variational Auto-Encoder designed for synthesizing tabular data, similar to CTGAN.  
    \item CODIGEM \cite{Walker2022} is a diffusion-inspired collaborative filtering model that utilizes stacking denoising autoencoders during the reverse denoising process.  
    \item MultiVAE \cite{liang2018variational} is a Variational Autoencoder that uses a multinomial likelihood loss objective and variational inference to model implicit user data.  
    \item MultiVAE++ is the pre-trained multinomial variational autoencoder used in SDRM, but its hyperparameters are optimized as shown in Table \ref{tab:search_space}. This model represents a true baseline to compare the increased performance of SDRM. 
    \item DiffRec \cite{Diffusion_Recommender_Model} is a denoising diffusion probabilistic model used for predicting whole user-item interactions.    
\end{itemize}

\subsubsection{Baseline Recommender Algorithms}

We evaluate the effectiveness of the synthetic data using three state-of-the-art recommendation algorithms: SVD (Singular Value Decomposition) \cite{mnih2007probabilistic}, a matrix factorization technique that approximates the original matrix with lower-dimensional matrices representing latent factors; MLP (Multilayer Perceptron) \cite{covington2016deep}, a deep neural network architecture incorporating user and item embeddings which feed into multiple feed-forward layers to model and predict whole user-item preferences; and NeuMF \cite{he2017neural}, that fuses matrix factorization and an MLP network to predict single user-item preferences. 

\subsubsection{Evaluation Metrics}

Similar to related work, we adopt two top-$k$ metrics, namely Recall@$k$ and NDCG@$k$. Recall@$k$ is defined as the percentage of ground-truth positives (items) that the system recommends as positive for a recommendation list of size $k$. The formula for Recall@$k$ is given as follows:

\begin{equation}
    \text{Recall@$k$}= \frac{\textup{Total Number of Relevant Items}}{\textup{Number of Relevant Items found in Top-k}}
\end{equation}

NDCG@$k$ (Normalized Discounted Cumulative Gain) is defined as the ratio of the cumulative gain to its ideal value, which measures how close the recommended top-$k$ ranking is to the ground truth. The formula for calculating NDCG at a specific rank position $k$ involves first computing the DCG (Discounted Cumulative Gain) at $k$, then normalizing that score by the IDCG (Ideal Discounted Cumulative Gain), which represents the maximum possible DCG at $k$ if the results were perfectly ranked. The formula is given as follows:

\begin{alignat}{2}
    & \text{DCG@k} && = \sum_{i=1}^{k} \frac{2^{rel_i} - 1}{\log_2(i + 1)} \\
    & \text{IDCG@k} && = \sum_{i=1}^{|REL_k|} \frac{2^{rel_i} - 1}{\log_2(i + 1)} \\
    & \text{NDCG@k} && = \frac{\text{DCG@k}}{\text{IDCG@k}}
\end{alignat}

Where $rel_{i}$ is the relevance score of the result at position $i$ and $\mid \text{REL}_k \mid$ is the list of relevance scores sorted in descending order up to position k.

\subsubsection{Generating Synthetic Data}
Generating data from CTGAN and TVAE was straightforward, as both approaches use a Gumbel-softmax \cite{gumble_softmax} to directly create categorical features. To generate data from CODIGEM, MultiVAE, MultiVAE++, and DiffRec, we first trained those models, then modified the code base to inject Gaussian noise into the decoder of MultiVAE and denoised Gaussian noise with CODIGEM and DiffRec. This process allows each baseline model to generate synthetic samples starting from Gaussian noise. 

For each synthetic dataset, we generated an equal number of samples to match the number of users in each training split of the original dataset. We utilized a conditional piece-wise function to transform the logit outputs from the generative baseline models and SDRM. We used a threshold value $\lambda$ on the synthetic data, which matches the value at the sparsity quantile of each dataset.

\begin{equation}
    f(x) = \begin{cases} 
1 & \text{if } x \geq \lambda \\
0 & \text{if } x < \lambda
\end{cases}
\label{eq:sparsity}
\end{equation}

\subsubsection{Hyperparameter Search for SDRM}

During the training procedure, we aim to find the set of hyperparameters \(x^{*}\) that yields the optimal \(f(x)\) where a hyperparameter $x$ can take any value in the search space of \(\chi\) \cite{yang2020hyperparameter}. We formalize the search as \(x^{*}=\arg\!\max_{x\epsilon \chi }f(x)\), where we optimize for Recall@10. We utilize a TPE (Tree-structured Parzen Estimator) \cite{TPE_sampler} and a successive halving pruner \cite{successive_halving_prunner} to search over the search space \(\chi\) efficiently. We perform 5-fold cross-validation for each set of hyperparameters 
used to train SDRM 
and report the average Recall and NDCG results. The hyperparameter search space used during SDRM training is shown in Table \ref{tab:search_space}.

\begin{table}[h]
    \centering
    \footnotesize
    \caption{Hyperparameter search space $\chi$ for SDRM}
    \begin{tabular}{|l|l|} \hline 
         \textbf{Hyperparameters}&  \textbf{Search Space}\\ \hline 
         SDRM epochs&  \{5, 501, step=5\}\\ \hline 
         SDRM learning rate&  \{1.e-6, 1.e-4, step=1.e-6\}\\ \hline 
         Gaussian variance&  \{0.01, 1, step=0.1\}\\ \hline 
 SDRM timesteps&\{3, 200, step=5\}\\ \hline 
 SDRM batch size&\{30, 1000, step=10\}\\ \hline 
         MLP hidden layers&  \{0, 5, step=1\}\\ \hline 
         MLP latent neurons&  \{50, 1000, step=50\}\\ \hline 
 MLP hidden layers& \{0, 5, step=1\}\\ \hline 
 VAE hidden layer neurons& \{MLP latent neurons, 1000, step=50\}\\ \hline 
 VAE learning rate: &\{1.e-4, 1.e-2, step=1.e-2\}\\ \hline 
 VAE batch size& \{30, 1000, step=10\}\\ \hline
    \end{tabular}
    \label{tab:search_space}
\end{table}

To find the optimal set of hyperparameters for SDRM, we employed Bayesian optimization search over the search space mentioned in Table \ref{tab:search_space} using the Optuna framework \cite{optuna2019}. Due to the observed variance in performance between runs with the same hyperparameters, we ran each set of candidate hyperparameters five times and averaged the results presented in Tables \ref{table:augmented_results} and \ref{table:only_synthetic_results}. We ran each evaluation model (SVD, MLP, NeuMF) with the benchmark datasets and SDRM between 30 to 300 trials. We computed these results using RTX 3080, RTX 4090, T4, V100, and A10 GPUs over several weeks. 

\subsection{Training Procedure}

We showcase the ability of diffusion to generate realistic user-item preferences in two conditions:  when the synthetic data is used to augment the original training set, and when the synthetic data is used to replace the original training set. For augmenting data, we combine the synthetic data with the training portion of the original dataset and evaluate the results on the holdout ratings from the test set. Similarly, for replacing the original data with synthetic, we train each baseline recommender algorithm with the synthetic data combined with 20\% of ground truth user ratings from the test set. This approach is necessary to show how synthetic data can enhance the predictive performance on real users, as it is impractical to measure the effectiveness of synthetic data in a recommendation model without incorporating some actual user data. During the evaluation process, we predict all possible items for each user in the test set and mask any items rated during training to prevent data leakage in our results. 

Each generative model was trained five times on each dataset and a synthetic dataset was generated matching the same number of users in its respective training set. Subsequently, each baseline recommender model was trained with and without the original data. Similarly, SDRM was trained five times and each run generated a synthetic dataset for the MultiVAE++ and SDRM using full (F-SDRM) and multi-resolution (M-SDRM) sampling. This way, every MultiVAE++, F-SDRM, and M-SDRM result was calculated from the same exact runs. We generate top-$k$ recommendations for each user in the validation set and compare them to the ground truth using Recall@$k$ and NDCG@$k$. 

To account for the negative sampling needed for training NeuMF, we utilized items with the lowest ratings from each user in the synthetic datasets. For the actual users, we selected the items with a 0 rating. 

\begin{table*}[h!]
\scriptsize
\caption{Overall performance comparison between baselines by training with synthetic and the original dataset. The best results are in bold and the second best are underlined. Average overall improvement: Recall 4.48\%, NDCG 5.07\%}
\scalebox{1}{\begin{tabular}{|lllllllll|}
\hline
\multicolumn{1}{|l|}{Model}       & \multicolumn{8}{c|}{SVD}                                                                                                                                                                                                                                                                               \\ \hline
\multicolumn{1}{|l|}{Dataset}     & \multicolumn{2}{c|}{ALB}                                                 & \multicolumn{2}{c|}{ML-100k}                                            & \multicolumn{2}{c|}{ML-1M}                                              & \multicolumn{2}{c|}{ADM}                                                \\ \hline
\multicolumn{1}{|l|}{Baseline}    & \multicolumn{1}{c}{Recall@10}       & \multicolumn{1}{c|}{NDCG@10}       & \multicolumn{1}{c}{Recall@10}      & \multicolumn{1}{c|}{NDCG@10}       & \multicolumn{1}{c}{Recall@10}      & \multicolumn{1}{c|}{NDCG@10}       & \multicolumn{1}{c}{Recall@10}      & \multicolumn{1}{c|}{NDCG@10}       \\ \hline
\multicolumn{1}{|l|}{Original}    & 0.3113 $\pm$  0                   & 0.287 $\pm$  0                   & 0.3716 $\pm$  0                  & 0.3973 $\pm$  0                  & \textbf{0.3769} $\pm$  0         & \textbf{0.4035} $\pm$  0         & \underline{0.0624} $\pm$  0      & \underline{0.0427} $\pm$  0      \\
\multicolumn{1}{|l|}{CTGAN}       & 0.2642 $\pm$  0.018              & 0.2407 $\pm$  0.014             & 0.3533 $\pm$  0.006              & 0.3813 $\pm$  0.004              & 0.3652 $\pm$  0.003             & 0.3917 $\pm$  0.004             & 0.0405 $\pm$  0.002             & 0.0267 $\pm$  0.001             \\
\multicolumn{1}{|l|}{TVAE}        & 0.3113 $\pm$  0                   & 0.2873 $\pm$  0                  & 0.3668 $\pm$  0.003             & 0.3918 $\pm$  0.003             & 0.3638 $\pm$  0.004             & 0.3869 $\pm$  0.003             & \underline{0.0624} $\pm$  0      & \underline{0.0427} $\pm$  0      \\
\multicolumn{1}{|l|}{CODIGEM}     & 0.3025 $\pm$  0.003              & 0.2819 $\pm$  0.002             & 0.3492 $\pm$  0.006             & 0.3745 $\pm$  0.01             & 0.3113 $\pm$  0.004             & 0.3336 $\pm$  0.004             & 0.049 $\pm$  0.003              & 0.0301 $\pm$  0.001             \\
\multicolumn{1}{|l|}{DiffRec}     & 0.2936 $\pm$  0.003              & 0.2748 $\pm$  0.003             & 0.37 $\pm$  0.008               & 0.4021 $\pm$  0.011             & 0.3598 $\pm$  0.007              & 0.383 $\pm$  0.008              & 0.0306 $\pm$  0.002             & 0.0228 $\pm$  0.003             \\
\multicolumn{1}{|l|}{MultiVAE}   & 0.3226 $\pm$  0.007              & \textbf{0.2998} $\pm$  0.007    & 0.3822 $\pm$  0.006             & 0.4103 $\pm$  0.007             & 0.3391 $\pm$  0.006             & 0.3658 $\pm$  0.007             & 0.0612 $\pm$  0.004             & 0.0389 $\pm$  0.002             \\
\multicolumn{1}{|l|}{MultiVAE++} & 0.3163 $\pm$  0.014             & 0.2912 $\pm$  0.007            & 0.3878 $\pm$  0.006             & 0.4126 $\pm$  0.007              & 0.3717 $\pm$  0.002             & 0.3946 $\pm$  0.003             & 0.0613 $\pm$  0.003             & 0.0406 $\pm$  0.002             \\ \hline
\multicolumn{1}{|l|}{F-SDRM}      & \textbf{0.325} $\pm$  0.012     & \underline{0.2988} $\pm$  0.01  & \underline{0.3924} $\pm$  0.005 & \textbf{0.4181} $\pm$  0.005     & \underline{0.3722} $\pm$  0.002 & \underline{0.3977} $\pm$  0.002  & 0.0623 $\pm$  0.003             & 0.041 $\pm$  0.001             \\
\multicolumn{1}{|l|}{M-SDRM}      & \underline{0.3249} $\pm$  0.004 & 0.2957 $\pm$  0.005            & \textbf{0.3971} $\pm$  0.004    & \underline{0.417} $\pm$  0.004  & 0.372 $\pm$  0.002               & 0.3962 $\pm$  0.003             & \textbf{0.0641} $\pm$  0.004    & \textbf{0.0442} $\pm$  0.002    \\ \hline
Improvement                       & \multicolumn{1}{c}{\textbf{0.74 \%}} & \multicolumn{1}{c}{\textbf{-0.33 \%}} & \multicolumn{1}{c}{\textbf{2.39 \%}} & \multicolumn{1}{c}{\textbf{1.33 \%}}  & \multicolumn{1}{c}{\textbf{-1.26 \%}} & \multicolumn{1}{c}{\textbf{-1.44 \%}} & \multicolumn{1}{c}{\textbf{2.72 \%}} & \multicolumn{1}{c|}{\textbf{3.51 \%}} \\ \hline
Model                               & \multicolumn{8}{c|}{MLP}                                                                                                                                                                                                                                                                               \\ \hline
\multicolumn{1}{|l|}{Original}    & 0.3183 $\pm$  0.006              & 0.3007 $\pm$  0.004             & 0.3569 $\pm$  0.007             & 0.3788 $\pm$  0.004              & 0.3319 $\pm$  0.006              & 0.3558 $\pm$  0.005             & 0.0194 $\pm$  0.001              & 0.0128 $\pm$  0.001             \\
\multicolumn{1}{|l|}{CTGAN}       & 0.3188 $\pm$  0.013              & 0.2987 $\pm$  0.005             & 0.3567 $\pm$  0.007             & 0.3781 $\pm$  0.006             & 0.3228 $\pm$  0.005             & 0.3463 $\pm$  0.005              & 0.0208 $\pm$  0.004             & 0.0135 $\pm$  0.002             \\
\multicolumn{1}{|l|}{TVAE}        & 0.2438 $\pm$  0.002              & 0.2431 $\pm$  0.001             & 0.3437 $\pm$  0.012             & 0.3622 $\pm$  0.014             & 0.2949 $\pm$  0.019             & 0.3186 $\pm$  0.02             & 0.0184 $\pm$  0.001             & 0.012 $\pm$  0.001              \\
\multicolumn{1}{|l|}{CODIGEM}     & \textbf{0.3365} $\pm$  0.014     & \textbf{0.3071} $\pm$  0.008    & 0.3479 $\pm$  0.01             & 0.3686 $\pm$  0.013             & 0.3132 $\pm$  0.007             & 0.3377 $\pm$  0.006             & 0.0596 $\pm$  0.003             & 0.0385 $\pm$  0.002             \\
\multicolumn{1}{|l|}{DiffRec}     & 0.3283 $\pm$  0.012              & 0.3013 $\pm$  0.004             & 0.355 $\pm$  0.011              & 0.3769 $\pm$  0.008             & 0.3246 $\pm$  0.001             & 0.3482 $\pm$  0.002             & 0.0213 $\pm$  0.002             & 0.0139 $\pm$  0.001             \\
\multicolumn{1}{|l|}{MultiVAE}   & 0.3328 $\pm$  0.009              & \underline{0.3048} $\pm$  0.006 & 0.053 $\pm$  0.014              & 0.0615 $\pm$  0.016             & 0.3588 $\pm$  0.003             & 0.372 $\pm$  0.006              & 0.0199 $\pm$  0.003             & 0.0137 $\pm$  0.001              \\
\multicolumn{1}{|l|}{MultiVAE++} & 0.3316 $\pm$  0.008              & 0.3 $\pm$  0.007                & \underline{0.3901} $\pm$  0.007 & \underline{0.4101} $\pm$  0.009 & 0.3528 $\pm$  0.005             & 0.3763 $\pm$  0.006             & \underline{0.0756} $\pm$  0.01 & \underline{0.0489} $\pm$  0.007 \\ \hline
\multicolumn{1}{|l|}{F-SDRM}      & \underline{0.3246} $\pm$  0.008  & 0.3004 $\pm$  0.004             & 0.3839 $\pm$  0.013             & 0.4055 $\pm$  0.014              & \textbf{0.3595} $\pm$  0.002    & \textbf{0.3845} $\pm$  0.001    & 0.0146 $\pm$  0.01             & 0.0095 $\pm$  0.008             \\
\multicolumn{1}{|l|}{M-SDRM}      & 0.3343 $\pm$  0.013              & 0.303 $\pm$  0.007              & \textbf{0.3947} $\pm$  0.011    & \textbf{0.419} $\pm$  0.01     & \underline{0.3591} $\pm$  0.003 & \underline{0.384} $\pm$  0.003  & \textbf{0.0798} $\pm$  0.0012    & \textbf{0.0523} $\pm$  0.002    \\ \hline
Improvement                       & \multicolumn{1}{c}{\textbf{-3.66 \%}} & \multicolumn{1}{c}{\textbf{-1.35 \%}} & \multicolumn{1}{c}{\textbf{1.18 \%}} & \multicolumn{1}{c}{\textbf{2.17 \%}}  & \multicolumn{1}{c}{\textbf{0.19 \%}} & \multicolumn{1}{c}{\textbf{2.18 \%}} & \multicolumn{1}{c}{\textbf{5.55 \%}} & \multicolumn{1}{c|}{\textbf{6.95 \%}} \\ \hline
Model                                 & \multicolumn{8}{c|}{NeuMF}                                                                                                                                                                                                                                                                             \\ \hline
\multicolumn{1}{|l|}{Original}    & 0.2399 $\pm$  0.009               & 0.0818 $\pm$  0.012             & 0.1006 $\pm$  0.009             & 0.096 $\pm$  0.005              & 0.0113 $\pm$  0.001             & 0.0094 $\pm$  0.001              & 0.0057 $\pm$  0.001             & 0.0011 $\pm$  0.001             \\
\multicolumn{1}{|l|}{CTGAN}       & 0.2444 $\pm$  0.002               & 0.0877 $\pm$  0.001             & 0.0889 $\pm$  0.012             & 0.086 $\pm$  0.013              & 0.0219 $\pm$  0.003             & 0.0192 $\pm$  0.004             & 0.007 $\pm$  0.002               & 0.0014 $\pm$  0.001             \\
\multicolumn{1}{|l|}{TVAE}        & 0.2439 $\pm$  0.001              & 0.0874 $\pm$  0.001             & 0.1258 $\pm$  0.011             & 0.1282 $\pm$  0.006             & 0.0126 $\pm$  0.004             & 0.0105 $\pm$  0.004             & 0.0053 $\pm$  0.001             & 0.0013 $\pm$  0.001             \\
\multicolumn{1}{|l|}{CODIGEM}     & 0.1709 $\pm$  0.116              & 0.0477 $\pm$  0.035             & 0.1015 $\pm$  0.016             & 0.0839 $\pm$  0.016             & 0.0642 $\pm$  0.001             & 0.0601 $\pm$  0.002             & 0.007 $\pm$  0.002               & 0.0014 $\pm$  0.001             \\
\multicolumn{1}{|l|}{DiffRec}     & 0.2621 $\pm$  0.006              & 0.0923 $\pm$  0.004             & 0.1147 $\pm$  0.013             & 0.1093 $\pm$  0.016             & 0.0203 $\pm$  0.005             & 0.0178 $\pm$  0.005             & 0.0138 $\pm$  0.002             & 0.0035 $\pm$  0.001             \\
\multicolumn{1}{|l|}{MultiVAE}   & 0.2475 $\pm$  0.057              & 0.0826 $\pm$  0.026             & 0.1253 $\pm$  0.012             & 0.118 $\pm$  0.014              & 0.0487 $\pm$  0.003             & 0.0483 $\pm$  0.002             & 0.0154 $\pm$ 0.004             & 0.0036 $\pm$ 0.001             \\
\multicolumn{1}{|l|}{MultiVAE++} & 0.2687 $\pm$  0.009              & 0.0877 $\pm$  0.01             & \underline{0.2357} $\pm$  0.003 & 0.1858 $\pm$  0.004             & \underline{0.1046} $\pm$  0.005  & \underline{0.0972} $\pm$  0.005 & 0.0224 $\pm$  0.007             & 0.005 $\pm$  0.002              \\ \hline
\multicolumn{1}{|l|}{F-SDRM}      & \textbf{0.3225} $\pm$  0.019     & \textbf{0.109} $\pm$  0.01     & 0.232 $\pm$  0.007              & \textbf{0.1891} $\pm$  0.009     & 0.1026 $\pm$  0.007             & 0.0935 $\pm$  0.007             & \underline{0.0234} $\pm$  0.003 & \underline{0.0054} $\pm$  0.001 \\
\multicolumn{1}{|l|}{M-SDRM}      & \underline{0.2953} $\pm$  0.019  & \underline{0.1026} $\pm$  0.006 & \textbf{0.2425} $\pm$  0.015    & \underline{0.186} $\pm$  0.008   & \textbf{0.1059} $\pm$  0.014    & \textbf{0.099} $\pm$  0.015     & \textbf{0.0273} $\pm$  0.0059    & \textbf{0.006} $\pm$  0.001     \\ \hline
Improvement                       & \multicolumn{1}{c}{\textbf{20.02 \%}} & \multicolumn{1}{c}{\textbf{24.28 \%}} & \multicolumn{1}{c}{\textbf{2.88 \%}} & \multicolumn{1}{c}{\textbf{1.77 \%}}  & \multicolumn{1}{c}{\textbf{1.24 \%}} & \multicolumn{1}{c}{\textbf{1.85 \%}} & \multicolumn{1}{c}{\textbf{21.87 \%}} & \multicolumn{1}{c|}{\textbf{20 \%}} \\ \hline
\end{tabular}}
\label{table:augmented_results}
\end{table*}

\begin{table*}[h!]
\scriptsize
\caption{Overall performance comparison between baselines by training with synthetic data. The best results are in bold and the second best are underlined. Average overall improvement: Recall 2.08\%, NDCG 0.88\% }
\scalebox{1}{
\begin{tabular}{|lllllllll|}
\hline
\multicolumn{1}{|l|}{Model}       & \multicolumn{8}{c|}{SVD}                                                                                                                                                                                                                                                \\ \hline
\multicolumn{1}{|l|}{Dataset}     & \multicolumn{2}{c|}{ALB}                                         & \multicolumn{2}{c|}{ML-100k}                                      & \multicolumn{2}{c|}{ML-1M}                                        & \multicolumn{2}{c|}{ADM}                                     \\ \hline
\multicolumn{1}{|l|}{Baseline}    & \multicolumn{1}{c}{Recall@10}   & \multicolumn{1}{c|}{NDCG@10}   & \multicolumn{1}{c}{Recall@10}   & \multicolumn{1}{c|}{NDCG@10}    & \multicolumn{1}{c}{Recall@10}   & \multicolumn{1}{c|}{NDCG@10}    & \multicolumn{1}{c}{Recall@10} & \multicolumn{1}{c|}{NDCG@10} \\ \hline
\multicolumn{1}{|l|}{Original}    & 0.3113 $\pm$ 0                  & 0.287 $\pm$ 0                  & 0.3716 $\pm$ 0                  & 0.3973 $\pm$ 0                  & \textbf{0.3769} $\pm$ 0         & \textbf{0.4035} $\pm$ 0         & \underline{0.0624} $\pm$ 0    & \underline{0.0427} $\pm$ 0   \\
\multicolumn{1}{|l|}{CTGAN}       & 0.0654 $\pm$ 0.054             & 0.056   $\pm$ 0.054            & 0.2215 $\pm$ 0.011             & 0.225 $\pm$ 0.012               & 0.2623 $\pm$ 0.008             & 0.2769 $\pm$ 0.008             & 0.0152   $\pm$ 0.002         & 0.0095   $\pm$ 0.001        \\
\multicolumn{1}{|l|}{TVAE}        & 0.2521  $\pm$ 0.005            & 0.2503   $\pm$ 0.004          & 0.2641 $\pm$ 0.008              & 0.2766 $\pm$ 0.006             & 0.3082  $\pm$ 0.006            & 0.325 $\pm$ 0.008              & 0.0214   $\pm$ 0              & 0.0143   $\pm$ 0             \\
\multicolumn{1}{|l|}{CODIGEM}     & 0.3044 $\pm$ 0.008             & 0.2813 $\pm$ 0.007            & 0.2799   $\pm$ 0.004           & 0.286 $\pm$ 0.007              & 0.2624  $\pm$ 0.005            & 0.2774 $\pm$ 0.005             & 0.0481   $\pm$ 0.003         & 0.0295   $\pm$ 0.001        \\
\multicolumn{1}{|l|}{DiffRec}     & 0.2702   $\pm$ 0.013           & 0.2541   $\pm$ 0.02          & 0.3476   $\pm$ 0.006           & 0.3624 $\pm$ 0.009             & 0.3354 $\pm$ 0.011             & 0.3595 $\pm$ 0.013             & 0.0258   $\pm$ 0.004          & 0.0197   $\pm$ 0.003        \\
\multicolumn{1}{|l|}{MultiVAE}   & 0.3299   $\pm$ 0.011           & 0.3063   $\pm$ 0.008           & 0.3724   $\pm$ 0.008           & 0.3937 $\pm$ 0.007             & 0.3134  $\pm$ 0.011             & 0.3359 $\pm$ 0.012             & 0.0393   $\pm$ 0.005         & 0.0246   $\pm$ 0.004        \\
\multicolumn{1}{|l|}{MultiVAE++} & \underline{0.3406} $\pm$ 0.003 & \underline{0.3188} $\pm$ 0.001 & 0.3889 $\pm$ 0.008             & 0.414 $\pm$ 0.008              & 0.3704 $\pm$ 0.004             & 0.3935 $\pm$ 0.004             &           0.0617 $\pm$ 0.003                    &           0.0412 $\pm$ 0.001                   \\ \hline
\multicolumn{1}{|l|}{F-SDRM}      & \textbf{0.3471}   $\pm$ 0.005  & \textbf{0.3229} $\pm$ 0.007   & \underline{0.3931} $\pm$ 0.006 & \textbf{0.4176} $\pm$ 0.003    & \underline{0.3707} $\pm$ 0.002 & 0.3944  $\pm$ 0.002            &        \textbf{0.0628} $\pm$ 0.002                       &        \textbf{0.0428} $\pm$ 0.001                      \\
\multicolumn{1}{|l|}{M-SDRM}      & 0.3384   $\pm$ 0.007           & 0.3174   $\pm$ 0.004          & \textbf{0.3946} $\pm$ 0.006    & \underline{0.4176} $\pm$ 0.006 & 0.3703  $\pm$ 0.003            & \underline{0.3946} $\pm$ 0.003 &      0.0622 $\pm$ 0.003                         &        0.0423 $\pm$ 0.002                      \\ \hline
Improvement                       & \multicolumn{1}{c}{\textbf{1.87\%}}      & \multicolumn{1}{c}{\textbf{1.27\%}}     & \multicolumn{1}{c}{\textbf{1.45\%}}      & \multicolumn{1}{c}{\textbf{0.86\%}}       & \multicolumn{1}{c}{\textbf{-1.65\%}}      & \multicolumn{1}{c}{\textbf{-2.21\%}}     & \multicolumn{1}{c}{\textbf{0.64\%}}          & \multicolumn{1}{c|}{\textbf{0.23\%}}        \\ \hline
Model       & \multicolumn{8}{c|}{MLP}                                                                                                                                                                                                                                                                    \\ \hline
\multicolumn{1}{|l|}{Original}    & 0.3183 $\pm$  0.006             & 0.3007 $\pm$  0.003             & \underline{0.3569} $\pm$  0.007 & \textbf{0.3788} $\pm$  0.004    & 0.3319 $\pm$  0.006             & 0.3558 $\pm$  0.005             & 0.0194 $\pm$  0.001              & 0.0128 $\pm$  0.001             \\
\multicolumn{1}{|l|}{CTGAN}       & 0.2651 $\pm$  0.008             & 0.2189 $\pm$  0.021             & 0.2017 $\pm$  0.016             & 0.2197 $\pm$  0.011             & 0.2081 $\pm$  0.001             & 0.2216 $\pm$  0.003             & 0.0011 $\pm$  0.001              & 0.0007 $\pm$  0.001             \\
\multicolumn{1}{|l|}{TVAE}        & 0.2438 $\pm$  0.001             & 0.2431 $\pm$  0.001             & 0.1334 $\pm$  0.007             & 0.1586 $\pm$  0.006             & 0.1569 $\pm$  0.017             & 0.1674 $\pm$  0.015             & 0.0153 $\pm$  0.003              & 0.0095 $\pm$  0.001             \\
\multicolumn{1}{|l|}{CODIGEM}     & 0.1568 $\pm$  0.095             & 0.1012 $\pm$  0.061             & 0.1445 $\pm$  0.019             & 0.1403 $\pm$  0.018             & 0.1882 $\pm$  0.006             & 0.187 $\pm$  0.01               & 0.0249 $\pm$  0.003              & 0.0173 $\pm$  0.001             \\
\multicolumn{1}{|l|}{DiffRec}     & 0.3065 $\pm$  0.005             & 0.2947 $\pm$  0.003             & 0.2997 $\pm$  0.016             & 0.3107 $\pm$  0.014             & 0.2349 $\pm$  0.033             & 0.2526 $\pm$  0.034             & 0.0181 $\pm$  0.002              & 0.0115 $\pm$  0.001             \\
\multicolumn{1}{|l|}{MultiVAE}   & 0.311 $\pm$  0.012              & 0.2925 $\pm$  0.012             & 0.3359 $\pm$  0.005             & 0.3465 $\pm$  0.007             & 0.3385 $\pm$  0.005             & 0.3633 $\pm$  0.005             & 0.0172 $\pm$  0.004              & 0.012 $\pm$  0.002              \\
\multicolumn{1}{|l|}{MultiVAE++} & 0.3356 $\pm$  0.011             & 0.3079 $\pm$  0.011             & 0.3488 $\pm$  0.01              & 0.3616 $\pm$  0.007             & 0.3409 $\pm$  0.004             & 0.3636 $\pm$  0.004             & \textbf{0.071} $\pm$  0.011      & \textbf{0.0492} $\pm$  0.007    \\ \hline
\multicolumn{1}{|l|}{F-SDRM}      & \underline{0.3375} $\pm$  0.008 & \underline{0.3103} $\pm$  0.01  & \textbf{0.3601} $\pm$  0.006    & 0.3754 $\pm$  0.006             & \textbf{0.3451} $\pm$  0.006    & \underline{0.3693} $\pm$  0.006 & 0.013 $\pm$  0.004              & 0.0085 $\pm$  0.003             \\
\multicolumn{1}{|l|}{M-SDRM}      & \textbf{0.3474} $\pm$  0.009   & \textbf{0.3142} $\pm$  0.005    & 0.3544 $\pm$  0.006             & \underline{0.3755} $\pm$  0.007 & \underline{0.3438} $\pm$  0.005 & \textbf{0.3702} $\pm$  0.004    & \underline{0.0668} $\pm$  0.005 & \underline{0.0451} $\pm$  0.004 \\ \hline
Improvement                       & \multicolumn{1}{c}{\textbf{3.51 \%}}  & \multicolumn{1}{c}{\textbf{2.04 \%}}  & \multicolumn{1}{c}{\textbf{0.89\%}}  & \multicolumn{1}{c}{\textbf{-0.87 \%}} & \multicolumn{1}{c}{\textbf{1.23\%}}  & \multicolumn{1}{c}{\textbf{1.81 \%}} & \multicolumn{1}{c}{\textbf{-6.28 \%}} & \multicolumn{1}{c|}{\textbf{-9.09\%}}  \\ \hline

Model                             & \multicolumn{8}{c|}{NeuMF}                                                                                                                                                                                                                                                                     \\ \hline
\multicolumn{1}{|l|}{Original}    & 0.2399 $\pm$  0.009             & 0.0818 $\pm$  0.011             & 0.1006 $\pm$  0.009             & 0.096 $\pm$  0.005              & 0.0113 $\pm$  0.001             & 0.0094 $\pm$  0.001             & 0.0057 $\pm$  0.001              & 0.0011 $\pm$  0.001             \\
\multicolumn{1}{|l|}{CTGAN}       & 0.1786 $\pm$  0.083             & 0.0405 $\pm$  0.018             & 0.0532 $\pm$  0.014             & 0.05 $\pm$  0.019               & 0.0053 $\pm$  0.001             & 0.0055 $\pm$  0.001             & 0.0073 $\pm$  0.002              & 0.0014 $\pm$  0.001             \\
\multicolumn{1}{|l|}{TVAE}        & 0.0957 $\pm$  0.034             & 0.0231 $\pm$  0.011             & 0.1289 $\pm$  0.011             & 0.1326 $\pm$  0.008             & 0.0237 $\pm$  0.005             & 0.023 $\pm$  0.005              & 0.009 $\pm$  0.008               & 0.0013 $\pm$  0.001             \\
\multicolumn{1}{|l|}{CODIGEM}     & 0.1154 $\pm$  0.108             & 0.0294 $\pm$  0.032             & 0.0782 $\pm$  0.014             & 0.0552 $\pm$  0.012             & 0.0606 $\pm$  0.006             & 0.0574 $\pm$  0.005             & 0.0022 $\pm$  0.001              & 0.0004 $\pm$  0.001             \\
\multicolumn{1}{|l|}{DiffRec}     & 0.2613 $\pm$  0.017             & 0.0881 $\pm$  0.006             & 0.0726 $\pm$  0.025             & 0.0629 $\pm$  0.027             & 0.0143 $\pm$  0.007             & 0.0134 $\pm$  0.006             & 0.0168 $\pm$  0.005              & 0.0031 $\pm$  0.001             \\
\multicolumn{1}{|l|}{MultiVAE}   & 0.0834 $\pm$  0.025             & 0.0196 $\pm$  0.007             & 0.0951 $\pm$  0.014             & 0.0894 $\pm$  0.014             & 0.0198 $\pm$  0.004             & 0.0195 $\pm$  0.004             & 0.0174 $\pm$  0.003              & 0.0042 $\pm$  0.001             \\
\multicolumn{1}{|l|}{MultiVAE++} & 0.331 $\pm$  0.01               & \textbf{0.1143} $\pm$  0.002    & \underline{0.2309} $\pm$  0.013 & \underline{0.1853} $\pm$  0.009 & 0.126 $\pm$  0.008              & \underline{0.1154} $\pm$  0.008 & 0.0259 $\pm$  0.004              & \underline{0.0059} $\pm$  0.001 \\ \hline
\multicolumn{1}{|l|}{F-SDRM}      & \textbf{0.3339} $\pm$  0.009    & 0.1137 $\pm$  0.002             & \textbf{0.2421} $\pm$  0.009    & \textbf{0.1913} $\pm$  0.007    & \underline{0.1265} $\pm$  0.011 & 0.1146 $\pm$  0.012             & \underline{0.028} $\pm$  0.014   & 0.0058 $\pm$  0.002             \\
\multicolumn{1}{|l|}{M-SDRM}      & \underline{0.332} $\pm$  0.008  & \underline{0.1141} $\pm$  0.001 & 0.2303 $\pm$  0.014             & 0.1814 $\pm$  0.01              & \textbf{0.1296} $\pm$  0.014    & \textbf{0.1212} $\pm$  0.014    & \textbf{0.0297} $\pm$  0.008     & \textbf{0.0064} $\pm$  0.002    \\ \hline
Improvement                       & \multicolumn{1}{c}{\textbf{0.87 \%}} & \multicolumn{1}{c}{\textbf{-0.17 \%}} & \multicolumn{1}{c}{\textbf{4.85 \%}} & \multicolumn{1}{c}{\textbf{3.23 \%}}  & \multicolumn{1}{c}{\textbf{2.85 \%}} & \multicolumn{1}{c}{\textbf{5.02 \%}} & \multicolumn{1}{c}{\textbf{14.67 \%}} & \multicolumn{1}{c|}{\textbf{8.47 \%}} \\ \hline
\end{tabular}}
\label{table:only_synthetic_results}
\end{table*}

\subsection{Results}
We compare SDRM against the baseline generative models used during SDRM training and present the average and standard deviation of Recall@$k$ and NDCG@$k$ over five runs in Tables \ref{table:augmented_results} and \ref{table:only_synthetic_results}. We also report the overall percentage improvement of SDRM over the best-performing baseline model for top-$10$ generated recommendations. In Table \ref{table:all_results}, we report the average improvement over MultiVAE++ in Recall@$k$ and NDCG@$k$, calculated over all the recommender algorithm and dataset combinations for both the augmented and the synthetic dataset training setups. We report additional improvements for various top-$k$ ($k \in {1, 3, 5, 10, 20, 50}$) over the MultiVAE++ baseline in the appendix in Tables \ref{tab:aug_improvement_over_multi} and \ref{tab:synthetic_improvement}.  

\subsubsection{Training with Augmented Dataset}
As shown in Table \ref{table:augmented_results}, F-SDRM and M-SDRM achieved the best or second-best results in all Recall@10 and all but one NDCG@10 results. There are many runs where MultiVAE++ placed second-best over either F-SDRM or M-SDRM, highlighting that even doing diffusion over a Gaussian latent variable can boost the performance. Overall, SDRM improves over the baseline generative methods by 4.48\% for Recall@10 and 5.07\% for NDCG@10 with the single best overall improvement from baseline of 24.28\% and 21.87\% for Recall and NDCG, respectively. 

\subsubsection{Replacing Training Data with the Synthetic Dataset}
As shown in Table \ref{table:only_synthetic_results}, when substituting the original datasets with synthetic datasets, SDRM again consistently achieves the best or second best overall results for each of the recommender models and averages an overall improvement of 2.08\% for Recall@10 and 0.88\% for NDCG@10, achieving as high as 14.67\% and 8.47\% for Recall and NDCG, respectively, on the Amazon Digital Music dataset. This demonstrates the potential of using SDRM as a method to substitute an original dataset and achieve competitive or improving performance. 

\subsubsection{Comparing SDRM against MultiVAE}
\begin{figure}[h]
    \centering
    \caption{Baseline improvement of SDRM over MultiVAE++ across all datasets for various sizes of top-$k$ recommendation lists. Augmented average improvement: Recall@$k$: 6.81\%, NDCG@$k$: 7.73\%, Synthetic average improvement: Recall@$k$: 1.79\%, NDCG@$k$: 1.56\%, Combined average improvement: Recall@$k$: 4.30\%, NDCG@$k$: 4.65\% }
    \includegraphics[width=8.5cm]{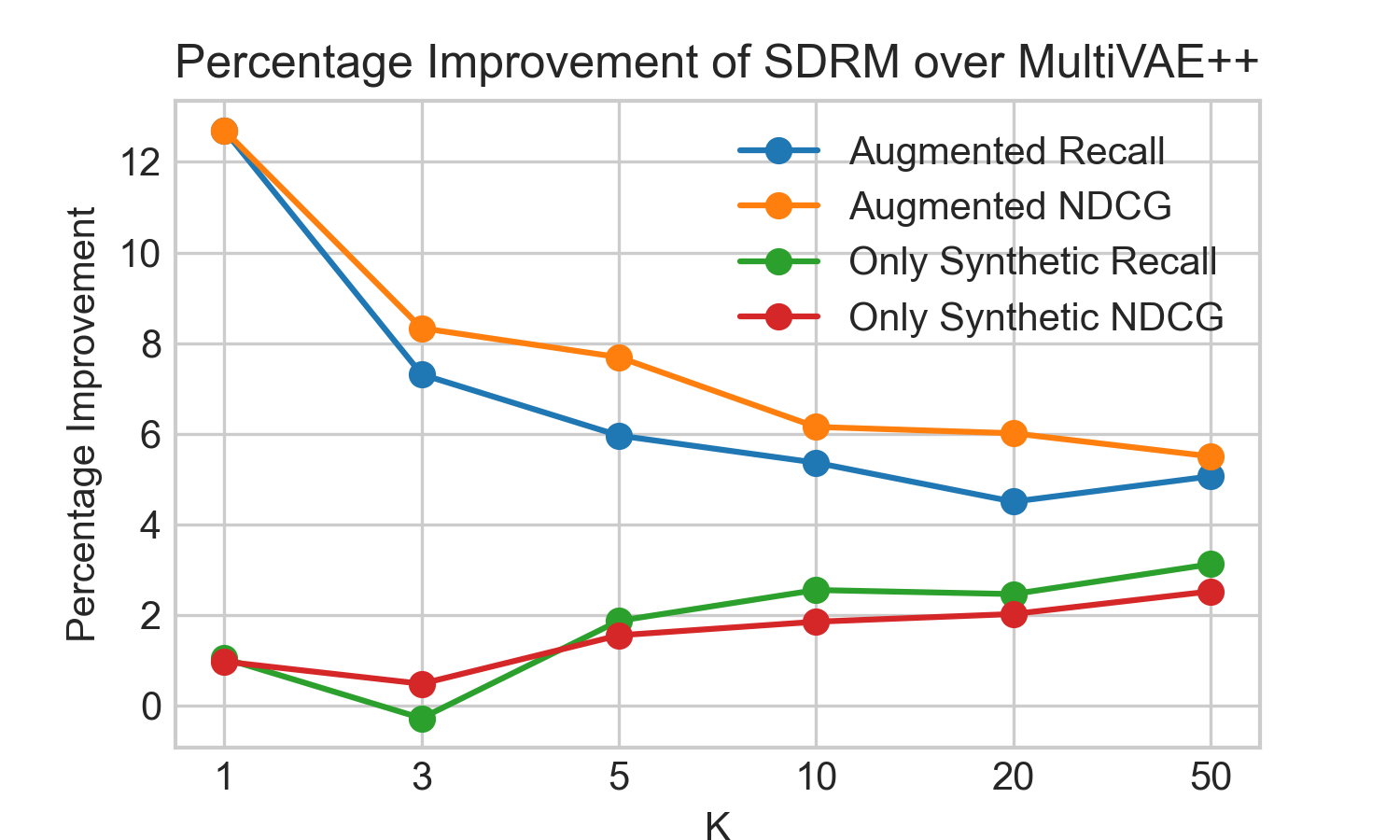}
    \vspace{-2em}
    \label{fig:percentage_improvement}
\end{figure}

Since SDRM employs a MultiVAE for compressing and representing user-item data as a smaller latent variable \(z\), then learns to transform one Gaussian representation into another, it raises a legitimate question whether the additional computational effort for SDRM genuinely enhances performance compared to using MultiVAE alone. To address this concern, we compared the percentage increase of max Recall and NDCG value between M-SDRM and F-SDRM (represented as SDRM in Figure \ref{fig:percentage_improvement}) against MultiVAE++ from the same runs. 

We observe that SDRM improves the overall Recall@$k$ and NDCG@$k$ scores compared to using MultiVAE across different $k$ values. Our method achieves an average 6.81\% and 7.73\% increase in Recall@$k$ and NDCG@$k$ when augmenting the original data with synthetic and 1.79\% and 1.56\% increase in Recall@$k$ and NDCG@$k$. The scores shown in Figure \ref{fig:percentage_improvement} and Table \ref{table:all_results} (located in the Appendix) are the average increase across each of the three baseline recommender models and four datasets for each Recall@$k$ and NDCG@$k$. SDRM has the greatest average performance increase in Recall for lower $k$  values scores when used for augmentation and a higher average performance increase in NDCG for larger $k$ scores when used as a substitution method. These findings further demonstrate that using a score-based diffusion model between the encoder and decoder of a MultiVAE architecture can significantly increase performance. 

\subsubsection{Discussion of Results}

We hypothesize that SDRM generates better synthetic data than MultiVAE++ because we use a score-based diffusion model to model the VAE's prior distribution $p_{\theta}(z)$. This allows us to create higher resolution noise processing before reconstruction using the decoder.

In the diffusion process, different timesteps \textit{t} affect data interpolation—larger \textit{t} values result in coarser and more varied interpolations, whereas smaller \textit{t} values preserve global features without fine details \cite{ho2020denoising, sohldickstein2015deep}. By sampling across a range of \textit{t} values, M-SDRM surpasses F-SDRM in capturing global patterns within latent \textit{z} space, which can improve recommender systems by better identifying intrinsic user or item clusters \cite{ma2019learning}.

Additionally, as shown by our extensive evaluation, our method surpasses other tabular data generation methods \cite{xu2019modeling}, previous recommendation variational autoencoders \cite{liang2018variational}, and earlier recommendation diffusion models \cite{Walker2022, Diffusion_Recommender_Model} in synthesizing recommendation datasets. We intentionally designed our approach to leverage the strengths of both variational autoencoders and diffusion models: the high-quality sampling capabilities of diffusion models and the mode coverage and diversity offered by variational autoencoders. This strategic combination enables us to outperform all existing data generation methods for recommendation datasets.

\subsection{Privacy-Preservation}

It has been previously shown that diffusion models and variational autoencoders can effectively reproduce training data even when sampled from a Gaussian distribution \cite{carlini2023extracting, Evaluating_Variational_Autoencoder}. It has also been observed that using a Gaussian distribution for generating synthetic data does not guarantee the privacy of the data \cite{heller2023can}. However, considering the high sparsity of recommendation data and the design of SDRM to generate binary item-rating preferences, we posit that our approach does not replicate the original training data, even when adhering to the same sparsity constraints as the original data.

To confirm that the proposed model is privacy-sensitive, we present empirical evidence that demonstrates that the overwhelming majority of SDRM-generated data does not reproduce the training examples from each dataset, while maintaining a similar distribution of rated items and user ratings as to the original. 

\subsubsection{Assessing Similarity}
To assess the difference between the synthetic and original data, we employ the Jaccard similarity metric, which quantifies the similarity between two sets by dividing the size of the intersection of the sets by the size of their union.

\begin{equation}
    J(x,\hat{x}) = {{|x \cap \hat{x}|}\over{|x \cup \hat{x}|}} = {{|x \cap \hat{x}|}\over{|x| + |\hat{x}| - |x \cap \hat{x}|}}
\end{equation}

\begin{figure}[h]
    \centering
    \caption{Distribution of the number of items and users for the Amazon Digital Music dataset.}
    \includegraphics[width=8.5cm]{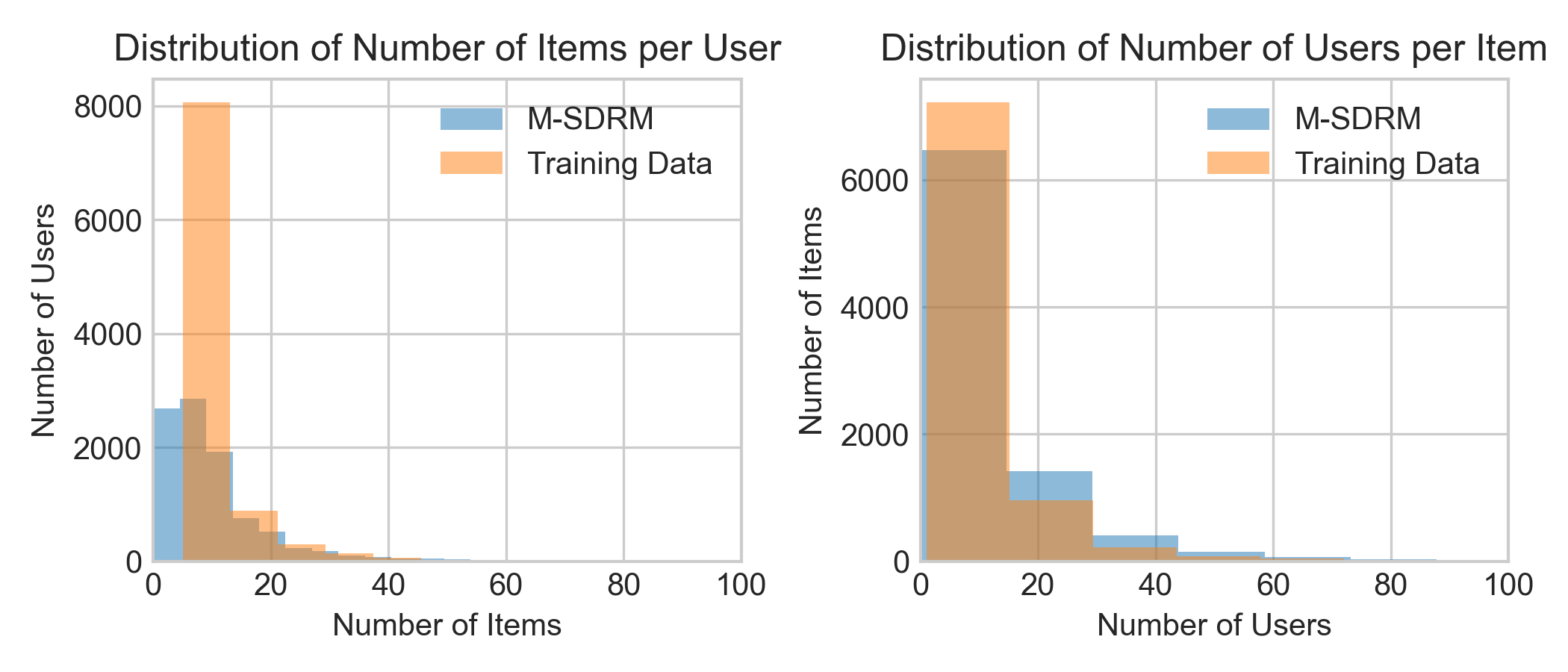}
    \vspace{-1em}
    \label{fig:adm_dist_users}
\end{figure}

We obtain a score for each pairwise comparison of each synthetic user \(\hat{x}\) to genuine user data  \(x\) using M-SDRM and the MultiVAE++ and present the results in Table \ref{tab:similarity_jacc}. We also explore the distribution of the items-per-user and user-per-items cardinalities in the synthetic data compared to the original dataset, as shown in Figure \ref{fig:adm_dist_users}. 

\begin{table}[h]
\caption{Pairwise comparison using Jaccard similarity on original data vs synthetic data with M-SDRM and MultiVAE++ on Amazon Digital Music dataset}
\small
\begin{tabular}{|l|ll|ll|}
\hline
Similarity & M-SDRM     & Total \% & MultiVAE++ & Total \% \\ \hline
0.0 - 0.1  & 90,331,367 & 98.88\%          & 89,949,754 & 98.46\%       \\
0.1 - 0.2  & 989,059    & 1.08\%           & 1,365,015  & 1.49\%        \\
0.2 - 0.3  & 32,636     & < 1\%            & 38,417     & < 1\%         \\
0.3 - 0.4  & 1,842      & < 1\%            & 1,796      & < 1\%         \\
0.4 - 0.5  & 320        & < 1\%            & 289        & < 1\%         \\
0.5 - 0.6  & 77         & < 1\%            & 64         & < 1\%         \\
0.6 - 0.7  & 43         & < 1\%            & 16         & < 1\%         \\
0.7 - 0.8  & 12         & < 1\%            & 8          & < 1\%         \\
0.8 - 0.9  & 8          & < 1\%            & 2          & < 1\%         \\
0.9 - 1.0  & 0          & < 1\%            & 3          & < 1\%         \\ \hline
\end{tabular}
\label{tab:similarity_jacc}
\end{table}

Even though each original dataset has been pre-processed to account for each user having a minimum of 5 ratings per item, SDRM produces synthetic users with less than five items (something that is not necessarily bad, given that it better reflects real datasets that have such cold-start users and items). Even so, SDRM still adheres to the natural long-tail distribution in recommendation data, where a smaller number of users (items) account for the majority of related item (user) ratings in the dataset, leading us to believe SDRM accurately models intricate item-rating distributions in a recommendation dataset. 

\subsubsection{Assessing Private Synthetic Data}
Table \ref{tab:similarity_jacc} reveals that using M-SDRM to generate datasets of the same size and sparsity as the original does not reproduce (copy) the training examples, resulting in almost 99\% dissimilarity. This significant difference is partly illustrated in Figure \ref{fig:adm_dist_users}, demonstrating that the synthetic dataset’s long-tail distribution of rated items is less right-skewed than the original. Despite this, the user-per-item distribution remains consistent between the synthetic and original datasets, highlighting SDRM’s effectiveness in synthesizing data while preserving preference distributions per item.

Overall, our analysis demonstrates that SDRM-generated data can balance the ability to protect users in privacy-sensitive circumstances while still offering the ability to train recommender system models with equal or better performance than the original, something that true differential privacy implementations inherently struggle with \cite{bagdasaryan2019differential}.

\section{Conclusion}
In this work, we propose SDRM, a novel approach to generating synthetic recommendation data through score-based diffusion modeling. We take advantage of the VAE framework to capture compressed latent representations into a Gaussian distribution and apply multi-resolution sampling to further improve distributional modeling for recommendation systems. Due to the disjointed training process and modeling of the latent {\it z} vector using score-based diffusion, we increase the performance of the variational inference process in a VAE when sampling new data, leading to significant improvements in performance in synthesizing recommendation datasets. Through various ablation studies, we show that our model is capable of modeling complex distributions and achieves an overall average improvement of 4.30\% in Recall@$k$ and 4.65\% in NDCG@$k$ when applying these novel samples to augment or replace existing datasets. We presented evidence that our method produces data that is 99\% dissimilar from the original dataset while still retaining the characteristics of the original data distribution. This work addresses the problem of dataset sparsity but also creates a novel way for organizations to preserve the privacy of their users' data while still training robust recommender systems. 

Our approach to synthesizing recommendation datasets using diffusion models addresses a gap in prior literature, which has primarily focused on synthesizing tabular datasets but often overlooks recommendation datasets in their benchmarks. We bridge this gap by not only comparing against popular tabular generators, but also evaluating recommender diffusion models and variational autoencoders.

\subsection{Limitations}

One limitation of our method is its reliance on a Multi-Layer Perceptron (MLP) network for denoising Gaussian data to generate user data one sample at a time. Unlike diffusion models used in image processing that utilize segmentation models for local pixel relationships, our approach may miss the chance to generate user data in batches, which could capture clustering patterns among users. Furthermore, our evaluations were limited to smaller recommendation datasets due to resource constraints. The largest dataset we tested, Amazon Digital Music, showcased SDRM's varying performance across different recommender models. This underscores the need for further research to evaluate SDRM's applicability and scalability to larger datasets. Additionally, our current method may not adequately model tabular datasets with continuous and categorical features of high cardinality, due to the heuristic approach used to enforce sparsity in SDRM's stochastic output.

\subsection{Future Work}

While we have demonstrated consistent improvements across various recommendation datasets, several avenues remain open for future research. Future efforts will aim to integrate auxiliary and temporal data to create synthetic data conditioned on user or group attributes. We believe this strategy can help mitigate biases and enhance fairness in recommendation datasets, enabling the training of fairness-aware recommendation models from a data-driven perspective. Further exploration could also yield success in experimenting with alternative noise perturbation techniques, network architectures, sampling methods, adversarial loss functions, and training methodologies.


\bibliographystyle{ACM-Reference-Format}
\bibliography{ref.bib}

\section{Additional Results Tables}

\begin{table*}[hbt!]
\small
\caption{Baseline improvement over MultiVAE++ across all datasets. Augmented average improvement: Recall@$k$: 6.81\%, NDCG@$k$: 7.73\%, Synthetic average improvement: Recall@$k$: 1.79\%, NDCG@$k$: 1.56\%, Combined average improvement: Recall@$k$ 4.30\%, NDCG@$k$ 4.65\% }
\scalebox{0.97}{\begin{tabular}{|l|llllllllllll|}
\hline
Metrics        & R@1      & R@3      & R@5     & R@10    & R@20    & R@50    & N@1      & N@3     & N@5     & N@10    & N@20    & N@50    \\ \hline
Augmented      & 12.7 \%  & 7.31 \%  & 5.96 \% & 5.36 \% & 4.5 \%  & 5.06 \% & 12.7 \%  & 8.33 \% & 7.69 \% & 6.15 \% & 6.01 \% & 5.50 \% \\
Only Synthetic & 1.04 \% & -0.28 \% & 1.87 \% & 2.55 \% & 2.46 \% & 3.12 \% & 0.97 \% & 0.48 \% & 1.55 \% & 1.85 \% & 2.02 \% & 2.52 \% \\ \hline
\end{tabular}}
\label{table:all_results}
\end{table*}

\begin{table*}[hbt!]
\scriptsize
    \centering
    \caption{SDRM augmented data percentage improvement over MultiVAE++}
    \begin{tabular}{|c|l|c|l|c|c|c|c|c|c|c|c|c|} \hline 
  Datasets &  \multicolumn{3}{|c|}{ML-100k}& \multicolumn{3}{|c|}{ALB}& \multicolumn{3}{|c|}{ML-1M}& \multicolumn{3}{|c|}{ADM} \\ \hline 
  Baseline Models &  SVD&NeuMF&MLP& SVD& NeuMF& MLP& SVD& NeuMF& MLP& SVD& NeuMF&MLP
 \\ \hline 
  Recall@1&  1.16
&18.81
&2.43& 6.04& 40.28& 5.29& 2.57& 7.92& 3.91& 5.00& 51.72&7.26
 \\ \hline 
          Recall@3&  0.45
&4.33
&3.57&  2.56&  9.62&  0.96&  1.37&  6.16&  2.99&  16.67&  30.95& 8.12
 \\ \hline 
          Recall@5&  0.29
&4.60
&1.68&  1.94&  13.89&  -0.47&  1.07&  1.76&  2.66&  16.20&  20.29& 7.64
 \\ \hline 
          Recall@10&  2.40
&2.89
&1.18&  1.86&  20.02&  0.81&  0.13&  1.24&  1.90&  4.57&  21.88& 5.56
 \\ \hline 
          Recall@20&  1.03
&-1.95
&1.76&  0.76&  18.38&  0.55&  0.40&  1.11&  2.10&  2.60&  20.11& 7.18
 \\ \hline 
          Recall@50&  0.84
&2.00
&0.10&  2.36&  17.60&  -0.05&  -0.02&  -1.08&  2.99&  5.15&  22.31& 8.61
 \\ \hline 
          NDCG@1&  1.16
&18.81
&2.43&  6.04&  40.28&  5.29&  2.57&  7.92&  3.91&  5.00&  51.72& 7.26
 \\ \hline 
          NDCG@3&  0.83
&7.06
&3.91&  3.43&  18.63&  1.47&  1.83&  6.80&  2.88&  16.07&  28.57& 8.49
 \\ \hline 
          NDCG@5&  0.70
&5.27
&2.65&  2.80&  22.31&  0.84&  1.49&  3.60&  2.64&  15.89&  25.93& 8.20
 \\ \hline 
          NDCG@10&  1.33
&1.78
&2.17&  2.68&  24.29&  1.00&  0.79&  1.85&  2.18&  8.87&  20.00& 6.95
 \\ \hline 
          NDCG@20&  0.62
&2.82
&1.83&  2.33&  22.52&  0.80&  0.71&  2.25&  2.18&  7.18&  21.43& 7.48
 \\ \hline
 NDCG@50& 0.82& 3.08& 1.12& 2.71& 21.20& 0.49& 0.42& 0.67& 2.49& 7.24& 17.65&8.12
 \\\hline
    \end{tabular}
    \label{tab:aug_improvement_over_multi}
\end{table*}

\begin{table*}[hbt!]
\scriptsize
    \centering
    \caption{SDRM synthetic training data percentage improvement over MultiVAE++}
\begin{tabular}{|c|cll|cll|cll|cll|}
\hline
\multicolumn{1}{|c|}{Datasets} & \multicolumn{3}{c|}{ML-100k}                                    & \multicolumn{3}{c|}{ALB}                                       & \multicolumn{3}{c|}{ML-1M}                                     & \multicolumn{3}{c|}{ADM}                                         \\ \hline
Baseline Models                & \multicolumn{1}{c|}{NeuMF} & \multicolumn{1}{l|}{MLP}   & SVD   & \multicolumn{1}{l|}{NeuMF} & \multicolumn{1}{l|}{MLP}  & SVD   & \multicolumn{1}{c|}{NeuMF} & \multicolumn{1}{l|}{MLP}  & SVD   & \multicolumn{1}{c|}{NeuMF}  & \multicolumn{1}{l|}{MLP}    & SVD  \\ \hline
Recall@1                       & \multicolumn{1}{c|}{-6.94} & \multicolumn{1}{l|}{13.36} & 2.42  & \multicolumn{1}{l|}{0.52}  & \multicolumn{1}{l|}{0.00} & 2.48  & \multicolumn{1}{c|}{9.88}  & \multicolumn{1}{l|}{3.90} & 1.30  & \multicolumn{1}{c|}{-15.11} & \multicolumn{1}{l|}{-9.06}  & 9.72 \\ \hline
Recall@3                       & \multicolumn{1}{c|}{2.07}  & \multicolumn{1}{l|}{6.40}  & -2.04 & \multicolumn{1}{l|}{-1.91} & \multicolumn{1}{l|}{2.05} & -2.01 & \multicolumn{1}{c|}{8.69}  & \multicolumn{1}{l|}{3.07} & 0.74  & \multicolumn{1}{c|}{-10.07} & \multicolumn{1}{l|}{-12.86} & 2.48 \\ \hline
Recall@5                       & \multicolumn{1}{c|}{3.73}  & \multicolumn{1}{l|}{5.23}  & 0.64  & \multicolumn{1}{l|}{-0.62} & \multicolumn{1}{l|}{2.26} & -0.65 & \multicolumn{1}{c|}{8.08}  & \multicolumn{1}{l|}{2.22} & 1.04  & \multicolumn{1}{c|}{4.74}   & \multicolumn{1}{l|}{-9.88}  & 5.67 \\ \hline
Recall@10                      & \multicolumn{1}{c|}{4.85}  & \multicolumn{1}{l|}{3.24}  & 1.47  & \multicolumn{1}{l|}{0.88}  & \multicolumn{1}{l|}{3.52} & 1.91  & \multicolumn{1}{c|}{2.86}  & \multicolumn{1}{l|}{1.23} & 0.08  & \multicolumn{1}{c|}{14.67}  & \multicolumn{1}{l|}{-5.92}  & 1.78 \\ \hline
Recall@20                      & \multicolumn{1}{c|}{2.31}  & \multicolumn{1}{l|}{-1.02} & 0.17  & \multicolumn{1}{l|}{-0.70} & \multicolumn{1}{l|}{1.72} & 1.76  & \multicolumn{1}{c|}{2.07}  & \multicolumn{1}{l|}{1.10} & 0.11  & \multicolumn{1}{c|}{31.18}  & \multicolumn{1}{l|}{-9.68}  & 0.52 \\ \hline
Recall@50                      & \multicolumn{1}{c|}{2.69}  & \multicolumn{1}{l|}{0.35}  & 1.37  & \multicolumn{1}{l|}{-1.87} & \multicolumn{1}{l|}{1.68} & 2.88  & \multicolumn{1}{c|}{1.03}  & \multicolumn{1}{l|}{0.23} & -0.02 & \multicolumn{1}{c|}{33.84}  & \multicolumn{1}{l|}{-6.02}  & 1.29 \\ \hline
NDCG@1                         & \multicolumn{1}{c|}{-6.94} & \multicolumn{1}{l|}{13.36} & 2.42  & \multicolumn{1}{l|}{0.52}  & \multicolumn{1}{l|}{0.00} & 2.48  & \multicolumn{1}{c|}{9.88}  & \multicolumn{1}{l|}{3.90} & 1.30  & \multicolumn{1}{c|}{-15.91} & \multicolumn{1}{l|}{-9.06}  & 9.72 \\ \hline
NDCG@3                         & \multicolumn{1}{c|}{0.64}  & \multicolumn{1}{l|}{7.92}  & -0.90 & \multicolumn{1}{l|}{-0.92} & \multicolumn{1}{l|}{1.44} & -0.39 & \multicolumn{1}{c|}{8.94}  & \multicolumn{1}{l|}{3.28} & 0.77  & \multicolumn{1}{c|}{-8.70}  & \multicolumn{1}{l|}{-11.97} & 5.59 \\ \hline
NDCG@5                         & \multicolumn{1}{c|}{2.08}  & \multicolumn{1}{l|}{5.47}  & 0.58  & \multicolumn{1}{l|}{0.00}  & \multicolumn{1}{l|}{1.48} & 0.19  & \multicolumn{1}{c|}{8.53}  & \multicolumn{1}{l|}{2.71} & 0.88  & \multicolumn{1}{c|}{1.30}   & \multicolumn{1}{l|}{-10.92} & 6.27 \\ \hline
NDCG@10                        & \multicolumn{1}{c|}{3.24}  & \multicolumn{1}{l|}{3.84}  & 0.87  & \multicolumn{1}{l|}{-0.17} & \multicolumn{1}{l|}{2.05} & 1.29  & \multicolumn{1}{c|}{5.03}  & \multicolumn{1}{l|}{1.82} & 0.28  & \multicolumn{1}{c|}{8.47}   & \multicolumn{1}{l|}{-8.33}  & 3.88 \\ \hline
NDCG@20                        & \multicolumn{1}{c|}{2.75}  & \multicolumn{1}{l|}{2.00}  & 0.27  & \multicolumn{1}{l|}{-0.65} & \multicolumn{1}{l|}{1.36} & 1.14  & \multicolumn{1}{c|}{3.79}  & \multicolumn{1}{l|}{1.55} & 0.27  & \multicolumn{1}{c|}{19.57}  & \multicolumn{1}{l|}{-9.80}  & 1.93 \\ \hline
NDCG@50                        & \multicolumn{1}{c|}{2.54}  & \multicolumn{1}{l|}{2.44}  & 0.73  & \multicolumn{1}{l|}{-0.90} & \multicolumn{1}{l|}{1.00} & 1.58  & \multicolumn{1}{c|}{2.82}  & \multicolumn{1}{l|}{1.05} & 0.26  & \multicolumn{1}{c|}{24.24}  & \multicolumn{1}{l|}{-8.10}  & 2.55 \\ \hline
    \end{tabular}
    \label{tab:synthetic_improvement}
\end{table*}

\end{document}